\newcommand{\nii}{[N~{\sc ii}]~$\lambda$6584~\AA}

\newcommand{\oiii}{[O~{\sc iii}]~$\lambda$5007~\AA}
\newcommand{\niii}{[Ni~{\sc ii}]~$\lambda\lambda$7378~\&~7412~\AA}
\newcommand{\kms}{${\rm km~s}^{-1}$}

\documentstyle{mn}
\title{The Manchester occulting mask imager (MOMI) -- first results 
on the
environment of P~Cygni}

\author[J.A. O'Connor, J. Meaburn and M. Bryce]
{J. A. O'Connor, J. Meaburn and M. Bryce\\ Department of Physics and
Astronomy, University of Manchester, Oxford Rd., Manchester, M13 9PL.}
\date{Accepted ??. 
      Received ??}
\begin{document}
\maketitle
\begin{abstract}
The design and first use of the Manchester occulting mask imager
(MOMI)
is described. This device, when combined with the Cassegrain or 
Ritchey--Chretien foci of 
large telescopes, is dedicated to the imagery of faint line emission
regions around bright central sources.

Initial observations, with MOMI on the Nordic Optical
telescope (NOT), of the V~$=$~4.8 mag P~Cygni environment, 
have revealed a 
$\geq$~5~arcmin long \nii\ emitting filament projecting from the
outer nebular shell of this luminous blue variable (LBV) star.
The presence of a mono--polar lobe older than both the 
inner (22~arcsec diameter) and outer (1.6~arcmin diameter) shells 
is suggested.
\end{abstract}
\begin{keywords}line: profiles - stars: individual: P~Cygni - stars:
mass-loss - ISM: bubbles.
\end{keywords}

\section{Introduction}

\noindent 
There are a variety of astrophysical problems which require both direct 
imagery and velocity imagery (i.e. images 
in small intervals of radial velocity)
of faint emission line regions in the close vicinity
of dominantly bright sources. For instance,
the nebulosities surrounding luminous blue variables (LBVs) are of
considerable interest for they are the relics of the most recent
eruptions of these stars. Expanding shells of circumstellar gas have
now been found around six galactic LBVs (see Barlow et al 1994 and
Nota et al 1995 for a summary of these observations).  So far, two
distinctly different shells have been found \cite{ba94} with
occulting--mask imagery around P~Cygni (V~=~4.8 mag). A bright inner
shell, $\approx$~ 22~arcsec diameter, has a radial expansion velocity
of 140~\kms\ in the \nii\ line but only 110~\kms\ in the exceptionally
bright \niii\ lines. A fainter, outer, \nii\ emitting shell of
$\approx$~1.6~arcmin diameter has been shown to be expanding at 185
\kms\ to give a kinematical age of 2100 yr for a distance of 1.8 kpc
to P~Cygni \cite{me96}.

The Manchester Echelle spectrometer, MES \cite{me84}, in its imaging
mode, and with an occulting strip in its focal plane, was used for
this initial imagery of P~Cygni. In this auxiliary mode, MES has a
very restricted field--of--view (1.9~arcmin~$\times$~1.5~arcmin) on
the Isaac Newton 2.5-m telescope consequently any nebular ejecta from
P~Cygni of larger angular diameter would remain undetected. The
Manchester occulting mask imager (MOMI) has now been manufactured to
overcome this restriction. This is a device dedicated to occulting
mask imagery and has had its first use on the Nordic Optical telescope
(NOT) where \nii\ images of the environs of P~Cygni have been obtained
over a field area of unprecedented size.

\section{Optical layout of MOMI}

\noindent The optical layout of MOMI, at the Ritchey--Chretien 
focus of the NOT telescope, is shown in Fig.~1. The light is
collimated for passage through a narrow--band interference filter,
centred on a nebular emission line and placed in the pupil. The field
is re--imaged on to the `science' CCD. A pressure stepped,
optically contacted, Fabry--Perot etalon can also be included
just before the filter in the pupil to permit the option 
of obtaining spatial/radial velocity
data  `cubes' of emission line regions
around dominantly bright central sources.
This option has not yet been used in the present instrument 
but was built into, and proven astrophysically, in
the forerunner \cite{me82} of MOMI.

The most critical aspect of the design is that the chromium occulting
mask, $\equiv$~4 arcsec on the sky, is on the first surface of the
optical chain. With this arrangement, when faint nebulosity around a
bright star (e.g. the V~$=$~4.8~mag P~Cygni) is being imaged, the only
contamination of the field is then by the starlight scattered in the
atmosphere and reflecting optics of the two--mirror telescope, along
with diffraction spikes caused by the spines of the telescope's
secondary mirror mounting. Any refracting optical component prior to
this mask floods the field with `ghosts' with such a bright star as a
result of multiple reflections. Incidentally, without a mask at all, 
the contamination by ghosts and scattering of all origins for P~Cygni
is unnacceptable with an integration time of only one second.
These contaminations are reduced by tens of thousands of times
with the use of the occulting mask. 

The optimum size of any mask must depend on the 
size of the maximum `seeing' disk encountered in one integration
yet be small enough not to occult useful information from any circumstellar
nebulosity. In this first use, a mask diameter of $\equiv$~4~arcsec
was thought to be a good compromise though other possibilities
were not explored.  

The chromium mask transmits 0.01 percent of the incident light at
6500\AA. This is useful in two ways. The position of the star relative
to the mask is recorded by the science CCD.  Also the stellar position
can be monitored throughout the integration by reflecting 8 percent of
this light out of the beam with a mylar beamsplitter and forming a
broad--band image on a secondary CCD, in this case a
thermo--electrically cooled Lynx. The transmission of the occulting
mask will increase over the 6500\AA\ value by 28 times at 8000\AA\
and decrease by 100 times at 5000\AA\ (Fr\'{e}edericks 1911).

For the present direct imaging 
observations at the f/11 Ritchey--Chretien focus of the
2.56-m NOT telescope the 710~mm focal length Tessar collimating lens
combined with the 300~mm focal length `off--the--shelf' Nikkor
refocussing lens gives a field scale of 1 arcsec~$=$~0.058~mm in the
MOMI focal plane.  With the thinned Loral 2048~$\times$~2048 CCD, 
which has
15~$\mu$m ($\equiv$~0.26~arcsec) square pixels, the field size is
8.83~arcmin~$\times$~8.83~arcmin.

\section{Observations and results}
Narrow band images of P~Cygni in
the light of \nii\ were obtained during the night of 1997 November
14 using the Loral CCD as the detector. A three--period (square
profile), 20~\AA\ bandwidth, interference filter was centred on the
\nii\ nebular emission line by tilting it by 3.2~deg. Fourteen
integrations, each of 500~s duration, 
were taken with P~Cygni centred on the
chromium spot.

The data were processed at the University of Manchester STARLINK node
with programs from the CCDPACK, FIGARO and KAPPA packages. The data
arrays were de-biassed and cosmic ray hits were
removed. Flat--field corrections were applied using 
dusk sky exposures as a reference. The resulting frames were aligned
and co--added to give an image with an effective integration time of
7000~s. A star of a similar brightness to P~Cygni was also observed in
a similar way to permit the correction of the residual scattered
stellar continuum from P Cygni.

The full field of MOMI was not used due to a significant drop in
quantum efficiency at two adjacent edges of the thinned Loral
chip. The occulting spot, which is mechanically offset by a small
amount from the centre of the chip, is further offset from the centre
of the reduced image when the affected edges are removed from the data
array.

A negative, high contrast, grey-scale representation of the trimmed
field (8.4~$\times$~8.4~arcmin$^{2}$) is shown in Fig.~2. An arc of
emission can be seen extending some 5~arcmin from the outer shell
towards the North--East of P~Cygni. In very
deep prints a marginally detected southern counterpart
to this 5~arcmin arc can be seen extending directly eastwards
from P~Cygni.

In Fig.~3 a subsection of the full field array enclosing the outer
shell is shown. Emission extending in the direction of the
aforementioned arc is visible. Small bow--shaped knots are apparent in
the outer shell at distances 47~arcsec , 48~arcsec , and 52~arcsec to
the North--East, South and South--West of the central star
respectively. The prominent dark feature to the North--West of the
central star is an artefact produced by ghosting within the layers of
the interference filter.

A positive grey-scale representation of the 
inner shell of the P~Cygni nebulosity \cite{ba94} 
is shown in Fig.~4. This is from a subsection of a single data
array with only a 500~s integration time. A two
dimensional analytical profile,
for radii greater than that of the occulting mask 
and with parameters taken from the
continuum reference data (for a star of similar brightness), 
has been subtracted from the P Cygni data in
this image. 
The central star is seen through the mask. Filamentary
loops and arcs within the inner nebular shell, 
particularly to the North--West of P~Cygni, are most
striking. Nebular 
knots are found throughout this shell. Diffraction spikes
from the spines of the secondary mirror are broadened due to the field
rotation corrector of the altazimuthly mounted NOT. The central
spike is due to a fine thread spanning the secondary mirror to secure
an alignment cap. Within the full co--added data array the rotation of
the altazimuth 
field causes unacceptable contamination to the image of the inner 
shell
by also rotating the diffraction spikes. 

\section{Discussion}
\subsection{Instrumental performance}
Narrow--band images of the nebulosity around the bright 
(V~=~4.8~mag) star P Cygni
have been obtained 
over a uniquely wide field 
with MOMI and with better resolution of
the small--scale features in the inner and outer nebular
shells than has been obtained previously. 
Furthermore, a faint, extensive, nebular arc has been discovered
apparently projecting from these two shells.
These observations were made
possible by the effectiveness of the chromium occulting 
mask at limiting the
amount of light, from the intense central star, which is transmitted
through the refracting optics. 
Consequently `ghosting' of transmitted starlight 
from the refractive elements has 
been reduced to a level that is
undetectable in the final image over the whole field.

The present version of MOMI has been designed to be transportable 
between several telescopes with focal ratios greater than
eleven
and it has not been optimised for one in  particular. 
Within this limitation, MOMI 
has performed to design expectations,
producing excellent imagery (0.9~arcsec FWHM in 0.7~arcsec seeing)
to a field radius of $\approx$\
4~arcmin. Off--axis optical aberrations 
in the corners of the field broaden
the stellar profiles to $\approx$\ 1.2~arcsec. The Tessar collimator
is operating beyond its design limit at the these extreme field angles
in a square f/11 field.

Secondary `ghosts' of the brightest parts of the nebulosity 
which originate within the multi--layers 
of the interference
filter (see Fig. 3) 
place restrictions on the filters that can be used with this
system. All multi--layer, narrow--band interference 
filters suffer from minor defects in their
construction. 
With careful selection of filters, and with 
tilt tuning of the passband position, the
filter `ghosts' can be minimised and placed away from areas of
interest. 
The filter `ghosts' could be removed by the
subtraction of an image of the same object with the filter
tilted by the same amount but in the opposite direction. 
However, for such a procedure to be successful,  
when MOMI is combined with an altazimuthly
mounted telescope such as the NOT, would also 
require careful alignment of the
diffraction spikes during observations and this was not attempted here.

The processing of data acquired with an equatorially mounted telescope
would be far simpler than that obtained with the 
altazimuthly mounted NOT. For instance, 
the fixed location of diffraction spikes would
facilitate their more complete removal during continuum subtraction. 
The spikes
themselves could also be reduced with the use of a fixed apodising mask
in the pupil near the interference filter (see Fig.~1) which is 
technically more complex to achieve 
when altazimuth field rotation is
involved.

The form of such an apodising mask must be tailored to the individual
telescope being used at any given time. 
The abrupt change in transmission over the narrow widths of the
secondary spines must be replaced by  graduated changes. A photographically
generated apodising mask, fixed in the pupil of an equatorial 
telescope, should reduce the intensity
of the diffraction spikes by a few times.
The apodising mask itself would have to rotate during the
integration time with respect to
the fixed pupil, for an altazimuth telescope, to counter the 
broadening of the diffraction spikes apparent  in Fig.~4.

\subsection{P Cygni phenomena}

The most striking feature of these initial observations of the
environment of P~Cygni is the
$\geq$~5~arcmin long faint nebular arc shown in Fig.~2.
This extensive arc appears to be associated with P~Cygni and not simply
foreground  or background nebulosity. It originates at the northern
edge of the outer nebular shell, it has a clear connecting
filament (see Fig.~3) to the latter and there is a hint of a complementary
southern arc, extending nearly west to east from P~Cygni. 
The
two arcs could then be the edges of a `mono-polar' lobe projecting
from P~Cygni. Obviously, the presence of this more southerly arc needs
confirming with much 
deeper imagery and a corresponding western 
lobe searched for. Single sided nebular lobes, projecting from stars, 
are are unknown whereas bipolar ones are common.  

With a distance to P~Cygni of 1.8 kpc \cite{ba94} then, if the
arc (or lobe) is nearly in the plane of the sky, its
linear extent is $\geq$~2.6~pc from P~Cygni. 
It is then comparable in dimensions to the extraordinary, faint
lobes projecting from the planetary nebula KjPn~8 \cite{lo95}.
Furthermore, its large angular extent suggests that its formation predates
both the 880 yr. age \cite{ba94} of the inner shell in Fig.~4  
and the 2100 yr. age \cite{me96} of the outer shell in Fig.3. 
Measurements of the kinematics of this extensive P~Cygni arc/lobe 
are now
required before speculating further about its age and origin.

It is interesting that the optical arc/lobe
shown in Fig. 2 appears to have a radio counterpart in the 
maps of Baars \& Wendker (1987) and Skinner et al (1998).
As a thermal origin for the radio emission is indicated,
radiative ionisation of this extensive feature seems likely.

\subsection{Broader use of MOMI}

The ionized environments close to
a large number of bright stars other than LBVs (e.g. Symbiotics,
O \& B stars etc.)
can be investigated with MOMI in its present form.
Additionally, chromium deposited occulting masks can easily be manufactured
to  
prescribed patterns and consequently, MOMI can be used 
beneficially on a broad range
of problems where central sources, of any shape, are extremely bright
compared with the adjacent line emission phenomena; 
e.g. where faint \oiii\ emitting 
halos of PNe surround dominantly bright
and compact (say 20~arcsec diam.), similarly \oiii\ emitting, cores.
The imagery of the Laques and Vidal (LV) knots \cite{la79} close to
the Trapezium group of stars would also benefit if all images
of the bright stars 
in the same field
were occulted simultaneously.

The pressure stepped, Fabry--Perot (see Sect. 2) 
option 
in the `science' arm (Fig.~1)
will have significant application for velocity imagery
which also requires an occulting mask. 
Investigation of the kinematics of the P~Cygni shells will require 
a Fabry--Perot of `finesse' thirty to give a spectral 
half-width $\equiv$~15 \kms\ at an
inter-order separation $\equiv$~450~\kms.  
This whole inter-order range can be covered
with the pressure of nitrogen as the scanning gas 
being increased to 5.36 atmos. above atmospheric pressure.
One particular merit of the pressure--stepped
Fabry--Perot is its operational simplicity.

\section{Acknowledgements}

JOC and JM acknowledge the excellent 
assistance of the staff at the Nordic Optical
telescope (NOT -- La Palma) where these initial observations were made.
In particular we are grateful to Hugo Schwartz for his assistance.
We thank the technical staff in our department for manufacturing
the device and PPARC for funding its construction. 
JOC also thanks PPARC for a Research Studentship.

\section{Legends}

\vspace{5mm}

\noindent {\bf Figure 1}\\
The optical layout of MOMI.

\vspace{5mm}

\noindent {\bf Figure 2}\\
The 8.4~$\times$~8.4~arcmin$^{2}$ image of the environs
of P~Cygni in the light of \nii.

\vspace{5mm}

\noindent {\bf Figure 3}\\
The outer shell of P Cygni in the light of \nii. The bright feature
to the NW of the star is a `ghost' generated within
the layers of the interference filter.

\vspace{5mm}

\noindent {\bf Figure 4}\\
The inner shell in the light of \nii. The residual scattered stellar
continuum has been subtracted by modelling that around a similarly
bright star. No attempt has been made to remove the prominent
diffraction spikes.

\bsp

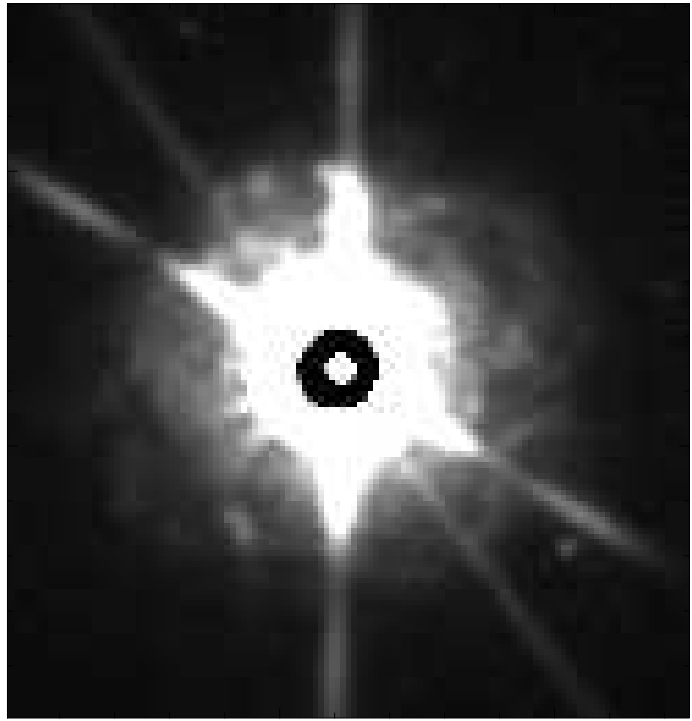
\end{document}